\documentclass{pasj00}

\SetRunningHead{Yamada \& Kitayama}{Infrared Emission from the ICD}
\title{Infrared Emission from Intracluster Dust Grains and 
Constraints on Dust Properties}
\author{Kenkichi \textsc{Yamada} and Tetsu \textsc{Kitayama}}
\affil{Department of Physics, Toho University,  
Funabashi, Chiba 274-8510, Japan}
\email{yamada@ph.sci.toho-u.ac.jp}
\Received{2005/03/31}
\Accepted{2005/06/06}
\begin{document}
\maketitle


\begin{abstract}

For 117 clusters of galaxies, we explore the detectability of
intracluster dust grains by current and future infrared facilities,
taking into account both collisional heating and sputtering of grains by
ambient plasma. If the dust grains are injected into the intergalactic
space with the amount and size comparable to the Galactic values, the
dust-to-gas ratio is typically $10^{-6}$ and the mean dust temperature
is $\sim 30$ K near the cluster center. The predicted infrared 
intensities lie marginally above the detection thresholds for Spitzer 
Space Telescope, ASTRO-F, Herschel and SPICA missions. For some nearby 
clusters such as Perseus, A3571, A2319, A3112 and A2204, good detections 
of intracluster dust signal are expected in the 70 $\mu$m band. Given 
rather tight constraints on the dust temperature from observed electron 
density and temperature, the dust mass can be inferred directly from the 
infrared observations. Further constraints on the size distribution will 
be obtained once multi-band data are available. They will definitely
provide a powerful probe of the dust injection processes and dust-gas
interactions in the intergalactic space.

\end{abstract}

\section{Introduction}

Dust grains have so far been detected only in the interstellar medium
and there is no firm detection in the intergalactic medium. The presence
of dust grains in the latter region is still an open question. If they
exist, the intergalactic dust grains should have great impacts not only
on our understanding of galaxy evolution but also on our interpretation
of high redshift observational data. From X-ray measurements of heavy
metal lines, it is evident that there is a significant amount of metal
outside of galaxies in clusters and groups of galaxies (e.g., Mushotzky
et al. 1996; Renzini 1997; Davis, Mulchaey \& Mushotzky 1999; Buote
2000). Such elements are likely to have been expelled out of galaxies
through supernova explosion, ram pressure stripping, or tidal
interaction. Dust could also be expelled by these processes or by some
independent mechanism such as radiation pressure (e.g., Chiao \& 
Wickramasinghe 1972; Ferrara et al. 1991; Shustov \& Vibe 1995; 
Davies et al. 1998; Simonsen \& Hannestad 1999; Aguirre et al. 2001a,b)

When dust grains are placed in the intracluster medium (hereafter ICM),
they would be heated by collisions with ambient hot X-ray emitting
electrons and ions, and emit mainly in infrared bands (Dwek et
al. 1990). At the same time, they are destructed via collisions with
impinging ions (sputtering) with a typical timescale of $\sim 10^8$
yr. The intracluster dust (hereafter ICD), if any, will therefore
provide a direct clue to the ejection processes of materials from
galaxies. Indeed, diffuse dust before destruction has been detected in 
a number of elliptical galaxies (e.g. Goodfrooij \& de Jong 1995). The
temperature, the amount, and the size distribution of the ICD, however,
can be quite different from those in the interstellar medium; in the
absence of stellar photons, temperature of the ICD will be lower than
that in elliptical galaxies. Moreover the ICD may be an important 
cooling agent in the ICM (Montier \& Giard 2004). Revealing the nature 
of the ICD will be of particular importance in understanding the 
dust-gas interaction.

There have been several suggestions and debates regarding the presence
of the intergalactic dust. For example, Girardi et al. (1992) suggested
that the redshift asymmetries of member galaxies in nearby groups are
consistent with the presence of dust in the intragroup medium. The
observed oxygen K edge in an X-ray spectrum of the Perseus cluster may
also be attributed to the ICD (Arnaud \& Mushotzky 1998). The extended 
submillimeter emission detected in a rich galaxy cluster may partly be 
due to dust (Komatsu et al. 1999; Kitayama et al. 2004). On the other 
hand, the presence of enhanced visual extinction towards high redshift 
objects behind clusters is still controversial (e.g., Maoz 1995).

More direct evidence has been searched for in the far-infrared bands.
Hickson et al. (1989) report that the far-infrared emission is enhanced
by a factor of 2 in compact groups of galaxies compared with a sample of
isolated galaxies. Sulentic \& De Mello Rabaca (1993), however, point
out that the results of Hickson et al. (1989) are likely to have been
overestimated mainly due to the limited spatial resolution of IRAS. The 
emission statistically detected in the IRAS data with many clusters of 
galaxies may partly be due to the ICD (Montier \& Giard 2005). 
Stickel et al. (1998 and 2002) observed 6 galaxy clusters
with ISOPHOT, and only in Coma they reported the excess of $\sim 0.2$
MJy/sr at 120 $\mu$m towards the central region. They estimated the
dust color temperature of $T_{\rm d} \sim 30$ K, the excess flux of
$F_{\rm ICD} \sim 2.8$ Jy and dust mass of $M_{\rm d} \sim 10^{7}
M_{\odot}$ from the ratio of surface brightness at 120 $\mu$m and 180 
$\mu$m. The visual extinction derived from the dust mass is a negligible 
amount ($A_{\rm V} \ll 0.1 $ mag). The result is in contrast to the 
reported optical extinction of $\sim 0.3$ mag for Coma (Zwicky 1962, 
Karachentsev \& Lipovetskii 1969).

The major difficulties in the past infrared observations are the limited
sensitivity and spatial resolution of the detector. The observational
data thus might be affected by Galactic cirrus emission, contamination
of individual galaxies, and confusion of extragalactic sources. The
infrared instruments on the current and future missions, such as Spitzer 
Space Telescope, ASTRO-F, Herschel, and SPICA, possess much improved
sensitivity and spatial resolution that are essential for separating the 
ICD emission from the other components. 

In this paper, we explore the possibility of detecting the ICD with 
Spitzer Space Telescope, ASTRO-F, Herschel and SPICA by performing 
a comprehensive study of the expected intensities from the ICD for 117 
clusters of galaxies at redshift $z \sim 0.01 - 0.8$. Based on currently 
available X-ray data, we explicitly take into account both heating and 
sputtering of dust grains via collisions with ambient hot plasma to 
compute their temperature and size distributions. We further examine the 
expected constraints from the near future observations on underlying dust 
model.

This paper is organized as follows. We describe our sample of galaxy 
clusters in Section \ref{sec:sample_clusters}. Details of the
model for the ICD size distribution is presented in Section
\ref{sec:model}. Section \ref{sec:results} describes the results for
the properties of the ICD and the expected infrared spectra, as well as
constraints from future observations on underlying dust model. We
discuss several uncertainties in the present analysis in Section
\ref{sec:discussion}. Finally, Section \ref{sec:conclusion} summarizes
our conclusions. Throughout the paper, we assume a standard set of
cosmological parameters; $\Omega_{\rm m} = 0.3$, $\Omega_{\Lambda} =
0.7$, $\Omega_{\rm b} = 0.04$, and $h = 0.7$. 


\section{Sample clusters}\label{sec:sample_clusters}

In order to calculate the ICD emission, we pick up 117 clusters at
redshift $z \sim 0.01 - 0.8$ for which gas properties can be determined
by X-ray spectroscopy and imaging data from ROSAT and ASCA. Unless
stated otherwise, we adopt an isothermal $\beta$-model for the gas
density profile of clusters.

We extract 79 clusters at $z \sim 0.1 - 0.8$ from Ota \& Mitsuda
(2004). We also supplement the sample with 38 nearby clusters at $z 
\sim 0.01 - 0.1$ from Mohr et al. (1999) and White (2000) for which 
the $\beta$-model parameters, temperature, and metalicity are all
available. For some clusters, Mohr et al. (1999) fit the X-ray
surface brightness by a sum of two $\beta$-model profiles.  In such
cases, we use the Abel's integral to deproject the fitted surface
brightness profile into the electron density profile (Yoshikawa \&
Suto 1999). For the three clusters (A1689, A2204 and PKS0745-19)
compiled in both Ota \& Mitsuda (2004) and Mohr et al. (1999), we use
the parameters from the former paper for definiteness.


\section{Model}\label{sec:model}

\subsection{Dust-gas interaction}

We first need to specify the basic dust model to calculate the ICD
emission. We consider a mixture of equal amounts of spherical silicate
and graphite as dust components, and adopt as their mass densities
$\rho_{\rm gra} = 2.26$ g cm$^{-3}$ (Draine \& Lee 1984) and $\rho_{\rm
sil} = 3.5$ g cm$^{-3}$ (Li \& Draine 2001), respectively. Using these
two components, the extinction curve in our Galaxy can be well
reproduced (Mathis, Rumpl \& Nordsieck 1977, hereafter MRN; Draine \&
Lee 1984; Laor \& Draine 1993).

If dust grains exist in the ICM, they will be heated by collision with
ambient X-ray emitting hot electrons and ions, and consequently 
emit in infrared band as they cool. The temperature of the ICD is then
obtained via the balance between collisional heating and radiative
cooling.

\begin{figure}
  \begin{center}
    \FigureFile(84mm,60mm){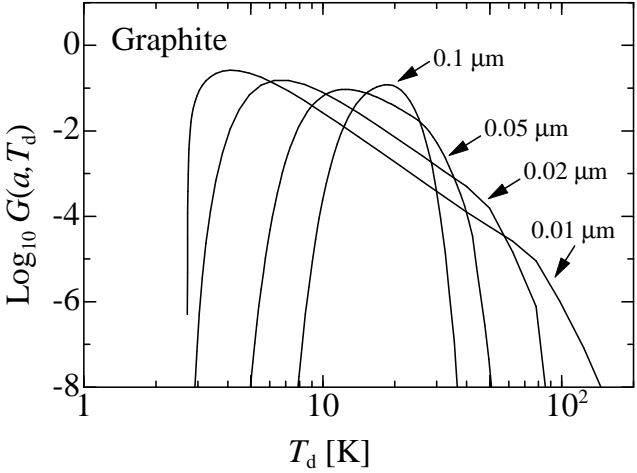}
  \end{center}
  \caption{Temperature distribution of graphite grains with various
 sizes embedded in a hot plasma with electron density $10^{-3}$
 cm$^{-3}$ and electron temperature $10^{8}$ K. } \label{fig:Td_dis}
\end{figure}

For given gas temperature and density, we determine the temperatures of
dust grains following Dwek (1986) taking account of both heating and
cooling between each collision. The heat capacities for graphite and
silicate are taken from Dwek (1986) and Draine \& Anderson (1985),
respectively. Figure \ref{fig:Td_dis} shows a specific example 
for the temperature distribution of graphite grains embedded in a
plasma with electron temperature $T_{\rm e} = 10^{8}$ K and electron
density $n_{\rm e} = 10^{-3}$ cm$^{-3}$. Note that the dust temperature
is not uniquely specified for given electron temperature and density,
but has a probability distribution depending on the grain size.  For
small ($a < 0.1 \mu$m) grains, the cooling time gets shorter than the
interval of collisions and the dust temperature has a large dispersion
(so-called stochastic heating). Some portion of these grains can have
rather high temperatures $\sim 100$ K owing to their small heat
capacity.  Larger grains, on the other hand, approach thermal
equilibrium with ambient plasma and the temperature distribution has a
sharp peak at $\sim 20$ K. As a result of stochastic heating, the
spectral shape of dust emission will be broadened particularly in the
Wien regime. We will explicitly take into account such features in our
prediction of the infrared emission from the ICD.

When dust grains are placed in a hot ICM, they are sputtered by
impinging ions.  We adopt the following analytic form 
for the sputtering rate (Tsai \& Mathews 1995):  
\begin{equation}
\frac{da}{dt} = - \tilde{h} \left(\frac{\rho_{\rm gas}}{m_{\rm p}}\right) 
\left[\left(\frac{T_{\rm s}}{T_{\rm gas}}\right)^{\omega} + 1\right]^{-1}, 
\label{sp_rate}
\end{equation}  
which gives a good approximation to the detailed calculations of Draine
\& Salpeter (1979) for both graphite and silicate grains when $\tilde{h}
= 3.2 \times 10^{-18}$ cm$^{4}$ s$^{-1}$, $T_{\rm s} = 2 \times 10^{6}$
K, and $\omega = 2.5$. Using this formula, we define the sputtering
timescale as
\begin{equation}
\tau_{\rm sputt}(a,r) \equiv a \mid da / dt \mid ^{-1}.
\end{equation}
For gas temperatures above $\sim 3 \times 10^{6}$ K, the sputtering
timescale is almost independent of gas temperature and approximated by
\begin{equation}
\tau_{\rm sputt}(a,r) \sim  10^{8} \  \left(\frac{a}{0.1 \mu \rm m}
\right) \left(\frac{n_{\rm e}}{10^{-3} {\rm cm}^{-3} }\right)^{-1}
~\rm yr.
\end{equation}

\subsection{Size distribution of dust grains in the ICM}

In modelling the size distribution of the ICD, we assume that the dust
grains are continuously injected into the ICM from galaxies and
destructed via sputtering. The size distribution $\partial
n_{i}(a;r)/\partial a$ for species $i$ at radial point $r$ then
follows (e.g., Dwek et al. 1990; Laor \& Draine 1993)
\begin{equation}
\frac{\partial \ }{\partial a}
\left(\dot{a}\frac{\partial n_{i}(a;r)}{\partial a}\right) + 
\frac{\partial}{\partial t}\frac{\partial n_{i}(a; r)}{\partial a} 
= \frac{\partial n_{i}^{\rm inj}(a;r)}{\partial a} 
\tau_{\rm cluster}^{-1},
\label{eq-master}
\end{equation}
where $\dot{a} = d a / d t$ is given by equation (\ref{sp_rate}),
$\partial n_{i}^{\rm inj}(a;r)/\partial a$ describes the total number
density of injected dust grains in the size interval between $a$ and 
$a + \delta a$ at radial point $r$, and $\tau_{\rm cluster}$ is the 
cluster lifetime over which dust grains have been injected. Neglecting
uncertainties for the formation epochs of individual galaxies in
clusters, we set $\tau_{\rm cluster}$ equal to the age of the Universe
at the cluster redshift.  For the injected dust grains, we assume that
they trace the galaxy distribution in a cluster and have the size
distribution observed in our Galaxy (MRN):
\begin{equation}
\frac{\partial n_{i}^{\rm inj}(a;r)}{\partial a} 
= A_{i}(r) a^{-\alpha^{\rm inj}} \ {\rm for} \ a_{\rm min}< a <a_{\rm max},
\end{equation}
where $\alpha^{\rm inj}=3.5$, $a_{\rm min} = 0.001 $ $\mu$m, $a_{\rm max} 
= 0.25$ $\mu$m, and $A_{i}(r) \propto \rho_{\rm gal}(r)$ is the
normalization coefficient to be determined later. Equation
(\ref{eq-master}) is solved for steady-state to give 
\begin{equation}
\frac{\partial n_{i}^{\rm std}(a;r)}{\partial a} 
= A_i(r) \frac{a^{-\alpha^{\rm inj}}}{\alpha^{\rm inj}-1} 
\frac{\tau_{\rm sputt}(a,r)}{\tau_{\rm cluster}}.
\label{eq-std}
\end{equation}
Since $\tau_{\rm sputt} \propto a$, the steady-state size distribution
is proportional to $a^{-\alpha^{\rm inj} + 1}$.

Given large uncertainties in the galaxy distribution in each cluster,
particularly at its outer envelope, we assume that galaxies trace
collisionless dark matter and $A_{i}(r) \propto \rho_{\rm gal}(r)
\propto \rho_{\rm\scriptscriptstyle DM}(r)$. We adopt the density
profile of dark matter inferred from N-body simulations (Navarro, Frenk
\& White 1996, hereafter NFW):
\begin{equation}
\rho_{{\rm\scriptscriptstyle DM}}(r) \propto 
 \frac{1}{(r/r_{\rm s})(1 + r/r_{\rm s})^{2}}
\end{equation}
where $r_{\rm s}$ is the scale radius of the halo. We follow Bullock et
al. (2001) to determine $r_{\rm s}$ for a halo with virial mass $M_{\rm
vir}$ at redshift $z$. As in Shimizu et al. (2003), we translate
iteratively the scaled mass $M_{500}$, given for each cluster in Mohr et
al. (1999) and Ota \& Mitsuda (2004), into $M_{\rm vir}$ using the
NFW profile, where $M_{500}$ is the mass enclosed within the radius at
which the mean density $\bar{\rho}(r)$ is equal to $500$ times the
critical density of the universe at the cluster redshift. The dark
matter density is normalized by $ M_{\rm vir} = \int_0^{r_{\rm vir}}
\rho_{\rm\scriptscriptstyle DM}(r) 4 \pi r^2 dr$.

Taken together, $A_i(r) $ is determined by fixing the total amount of
the injected dust grains $M_{\rm d} ^{\rm inj}$ as
\begin{eqnarray}
 M_{\rm d} ^{\rm inj}(<r_{\rm vir}) &=& \sum_{i}\frac{ 4\pi \rho_i}{3}   
\int_{0}^{r_{\rm vir}} 4 \pi r^2 A_{i}(r) dr    
\int_{a_{\rm min}}^{a_{\rm max}} a^{3-\alpha^{\rm inj}} da 
\nonumber \\
&=& Z_{\rm d}^{\rm inj} Z_{\rm metal} M_{\rm gas}(<r_{\rm vir}),
\end{eqnarray}
where the injected dust-to-gas mass ratio $Z_{\rm
d }^{\rm inj}$ is fiducially fixed at 
$0.0075$ (MRN), and the mean metalicity $Z_{\rm metal}$
in units of the solar value is taken from Ota \& Mitsuda (2004) and
White et al. (2000).  We have assumed that $Z_{\rm metal} M_{\rm
gas}(<r_{\rm vir})$ denotes the cumulative mass of the gas that has been
supplied from galaxies into the ICM. Using the assumed relation of
$A_i(r) \propto \rho_{\rm\scriptscriptstyle DM}(r)$, we obtain
\begin{equation}
A_i(r) = \frac{3 f_i}{4\pi \rho_i} 
\frac{Z_{\rm d}^{\rm inj} Z_{\rm metal}}{\int_{a_{\rm min}}^{a_{\rm max}} 
a^{3-\alpha^{\rm inj}} da } \frac{M_{\rm gas}(<r_{\rm vir})}{M_{\rm vir}} 
\rho_{\rm\scriptscriptstyle DM}(r),  
\end{equation}
where $f_{i}$ represents the mass fraction of graphite or silicate
grains.

\subsection{Infrared emission from the ICD}

Given the size distribution (eq.~[\ref{eq-std}]) and the
temperature distribution $G_{i}(a,T_{\rm d})$ of the ICD, the infrared
intensity at wavelength $\lambda$ is given by an integral over the 
line-of-sight through the cluster:
\begin{eqnarray}
I_{\lambda}(R) 
&=& \sum_{i} 
\int_{-l_{\rm max}}^{+l_{\rm max}} dl 
\left[\int_{a_{\rm min}}^{a_{\rm max}} da ~ 
\pi a^{2} Q_{\lambda}^{i}(a) \frac{\partial n_{i}^{\rm std}(a;r)}
{\partial a}  \right. \nonumber\\
&& \times \left.\left\{\int_{T_{\rm min}}^{T_{\rm max}} 
G_{i}(a,T_{\rm d}) B_{\lambda}(T_{\rm d})dT_{\rm d}\right\}\right], 
\end{eqnarray}
where $Q_{\lambda}^{i}(a)$ is the dust absorption efficiency factor of
dust species $i$ taken from Draine \& Lee (1984), $B_{\lambda}$ is the
Planck function, $R$ is the projected separation from the cluster
center, and $l$ is the length over a line of sight with $l_{\rm max} =
\sqrt{r_{\rm vir}^2 - R^2}$. The minimum temperature of the dust
$T_{\rm min}$ is set equal to the CMB temperature $2.73(1+z)$ K, and the
maximum temperature $T_{\rm max}$ to the vaporization temperature, 2000
K for graphite and 1500 K for silicate, respectively (Dwek 1986). As
$G_{i}(a,T_{\rm d})$ declines rapidly toward high temperatures, our
results are insensitive to a specific choice of $T_{\rm max}$.

\section{Results}
\label{sec:results}

\subsection{Properties of the ICD}

Based on the model constructed in the previous section, the ICD
mass density as a function of physical radius $r$ from the cluster
center is given by
\begin{equation}
\rho_{\rm d}(r) = \sum_{i} f_{i} \int_{a_{\rm min}}^{a_{\rm max}} 
\frac{4\pi a^{3} \rho_{i} }{3} 
\frac{\partial n_{i}^{\rm std}(a;r)}{\partial a} da. 
\end{equation}
The dust-to-gas mass ratio in the ICM can then be written as
\begin{equation}
Z_{\rm d}(r) = \frac{\rho_{\rm d}(r)}{\rho_{\rm gas}(r)}.
\end{equation}
We define the mass weighted mean temperature and mean size of the
ICD as follows:
\begin{eqnarray}
\langle T_{\rm d}(r)\rangle 
&=& \left[ \sum_{i} f_{i} \int_{a_{\rm min}}^{a_{\rm max}} 
\left\{
\int_{T_{\rm min}}^{T_{\rm max}} G_i (a,T_{\rm d}) T_{\rm d} 
dT_{\rm d} \right\} \right. 
\nonumber  \\ 
&&\times  \left. \left. \frac{4\pi a^{3} \rho_{i}}{3} 
\frac{\partial n_{i}^{\rm std}(a;r)}{\partial a} da 
\right] \right/ \rho_{\rm d}(r),  \\
\langle a_{\rm d}(r)\rangle 
&=& \left[ \sum_{i} f_{i} \int_{a_{\rm min}}^{a_{\rm max}} 
\frac{4\pi a^{3} \rho_{i} }{3} \right.
\nonumber \\
&&\hspace{10mm} \times  \left. \left. 
\frac{\partial n_{i}^{\rm std}(a;r)}{\partial a} a da 
\right] \right/  \rho_{\rm d}(r). 
\end{eqnarray}
For later convenience, we further introduce 
the surface mass density of the ICD at a projected separation $R$ from
the cluster center by 
\begin{equation}
\Sigma_{\rm d}(R) = \int_{-l_{\rm max}}^{+l_{\rm max}} 
\rho_{\rm d}(r) dl.   
\end{equation}
  
\begin{figure}
  \begin{center}
    \FigureFile(84mm,60mm){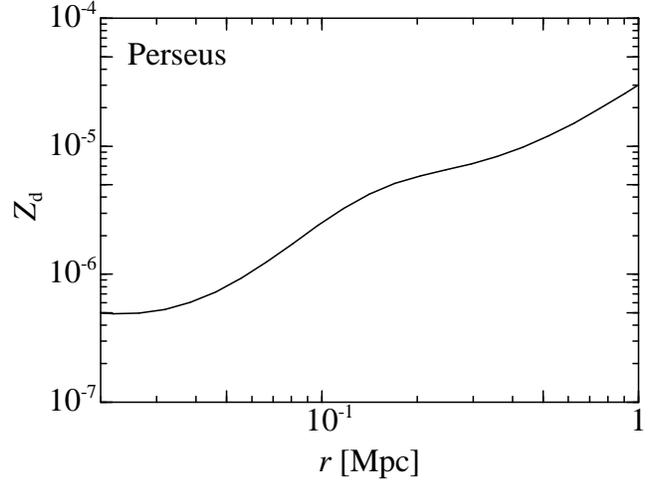}
  \end{center}
  \caption{Dust-to-gas mass ratio in the ICM of Perseus cluster (A426) 
  as a function of physical radius from the center.}
  \label{fig:Per_Zd}
\end{figure}

Figures \ref{fig:Per_Zd}--\ref{fig:Per_Sd} show the above
quantities in a specific case of Perseus cluster (A426).  As will be
shown below, the expected intensity from this cluster is among the
highest in our entire sample and we will take it as a representative
target in our subsequent analysis. The dust-to-gas mass ratio $Z_{\rm
d}(r)$ is lower, particularly near the center, by several orders of
magnitude than the Galactic value $0.0075$ as a result of rapid
sputtering (Fig. \ref{fig:Per_Zd}). The mean dust temperature $\langle
T_{\rm d}(r)\rangle$, on the other hand, is the highest near the center
at $> 30$ K due to efficient heating by electron collisions
(Fig. \ref{fig:Per_Tdave}).  The mean size of the dust is independent of
$r$ and its value in Perseus is $\sim 0.15$ $\mu$m. Once integrated
over the line-of-sight, the ICD surface mass density $\Sigma_{\rm d}$
remains nearly constant within the projected separation of $\sim
300$ kpc (Fig. \ref{fig:Per_Sd}) and will serve as a good measure of
the total amount of the ICD.

\begin{figure}
  \begin{center}
    \FigureFile(84mm,60mm){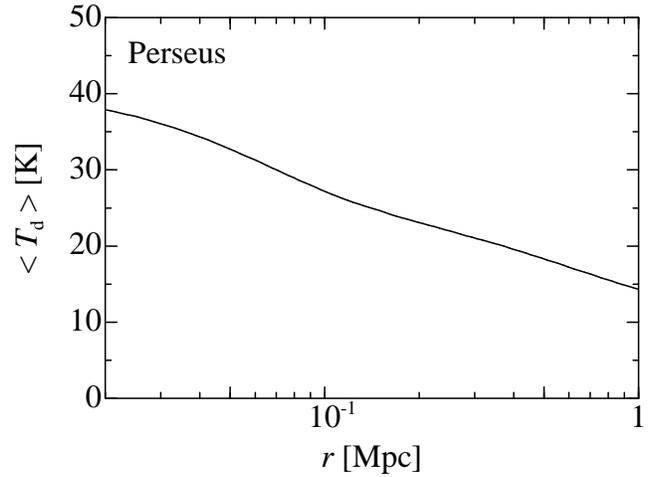}
  \end{center}
  \caption{Mass weighted mean temperature of the ICD in Perseus as a
  function of physical radius from the center.}  \label{fig:Per_Tdave}
\end{figure}

\begin{figure}
  \begin{center}
    \FigureFile(84mm,60mm){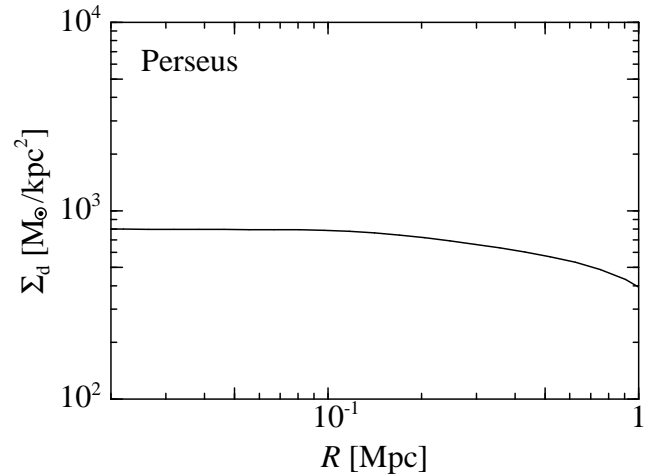}
  \end{center}
  \caption{Surface mass density of the ICD in Perseus as a function
  of projected separation from the cluster center.}
  \label{fig:Per_Sd}
\end{figure}

\subsection{Expected infrared spectra}

\begin{figure}
  \begin{center}
    \FigureFile(84mm,60mm){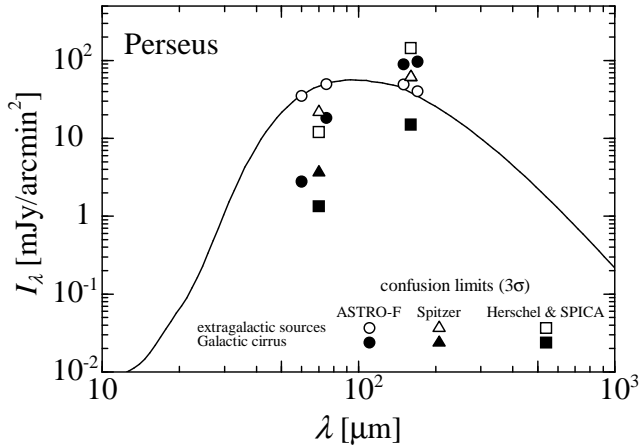}
  \end{center}
  \caption{Predicted spectra of the ICD at the projected separation
  of 20 kpc from the center of Perseus cluster.
  Open circles, open triangles and open squares show the 3$\sigma$
  confusion limits, due to extragalactic sources, for ASTRO-F, Spitzer,
  Herschel and SPICA missions, respectively. Filled symbols with an
  identical shape indicate those due to Galactic cirrus for the same
  instrument. Note that the open and filled triangles overlap each
  other at 160 $\mu$m.} 
  \label{fig:Per_Int}
\end{figure}

We plot the predicted ICD emission spectra from Perseus cluster in
Figure \ref{fig:Per_Int}. In the rest of the paper, unless otherwise
stated explicitly, the emission is computed at the projected separation
of 20 kpc from the center in order to avoid any contamination from the
central galaxy. Also plotted for reference are the $3\sigma$ confusion 
limits due to extragalactic sources, for Spitzer/MIPS (Dole et al. 2004),
ASTRO-F/FIS (Pearson et al. 2004), Herschel and SPICA (Dole et
al. 2004), as well as those due to Galactic cirrus (Jeong et al. 2005).
The confusion levels are computed at the beam size of each
instrument. For simplicity, we represent band filters of FIS (N60,
N170, WIDE-S and WIDE-L) by their central wavelengths 60, 170, 75 and
150 $\mu$m, respectively. We adopt 70 $\mu$m and 160 $\mu$m for the
band filters of Herschel and SPICA. The PSF of each instrument is
modeled by a Gaussian with the FWHM value taken from the reference
mentioned above. We also take into account the dependence of the
cirrus emission on the Galactic latitude as in Jeong et al. (2005).
The total confusion limit should be estimated as $\sigma_{\rm tot} =
\sqrt{\sigma_{\rm extragalactic}^{2}+\sigma_{\rm cirrus}^{2}}$. This
figure indicates that the source confusion is dominated by extragalactic
sources at $\lambda < 100\mu$m, while the contribution of Galactic cirrus
increases rapidly at longer wavelengths. The predicted emission has its
peak at $\sim 70 - 100$ $\mu$m and would be detectable above $5\sigma$
levels at 70 $\mu$m with Spitzer, Herschel and SPICA, and marginally
detectable at the $3\sigma$ levels at 60 and 75 $\mu$m with ASTRO-F.

\begin{figure}
  \begin{center}
    \FigureFile(84mm,60mm){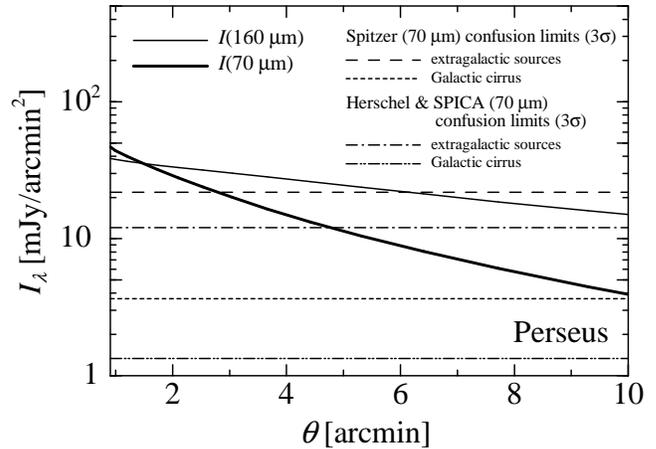}
  \end{center}
  \caption{Expected radial profiles of the ICD emission at 70 $\mu$m
  (thick solid line) and 160 $\mu$m (thin solid line) in Perseus
  cluster. Also plotted for reference are the 3$\sigma$ confusion
  limits at 70 $\mu$m; extragalactic sources for Spitzer (dashed line),
  Herschel and SPICA (dash-dotted line), and Galactic cirrus for Spitzer
  (dotted line), Herschel and SPICA (dash-dot-dotted line).}
  \label{fig:Per_radipro}
\end{figure}

Figure \ref{fig:Per_radipro} shows the expected radial profiles for the
emission from Perseus. The emission at 70 $\mu$m would be detectable
above the $3\sigma$ levels within $\sim$ 2.5 arcmin by Spitzer 
and within $\sim$ 4.5 arcmin by Herschel and SPICA.
At the outer envelopes of the cluster, the emission declines owing to the 
lack of hot gas capable of heating the dust. The extended feature of the
emission will be of particular importance in separating the ICD
component from galaxies. A fine spatial resolution of Spitzer, ASTRO-F,  
Herschel and SPICA will be essential for this purpose. 

\begin{figure*}
  \begin{center}
    \FigureFile(84mm,60mm){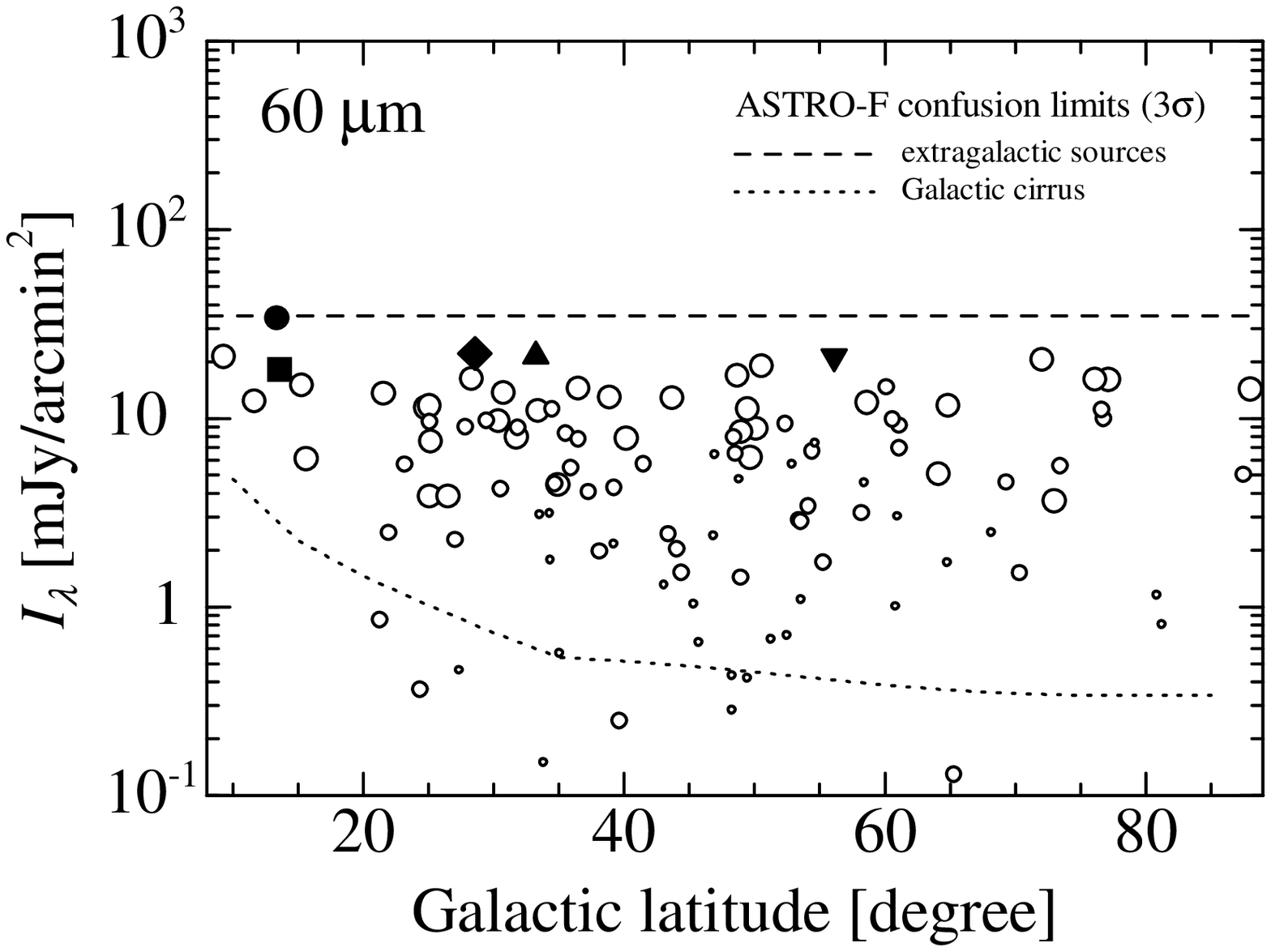}
    \FigureFile(84mm,60mm){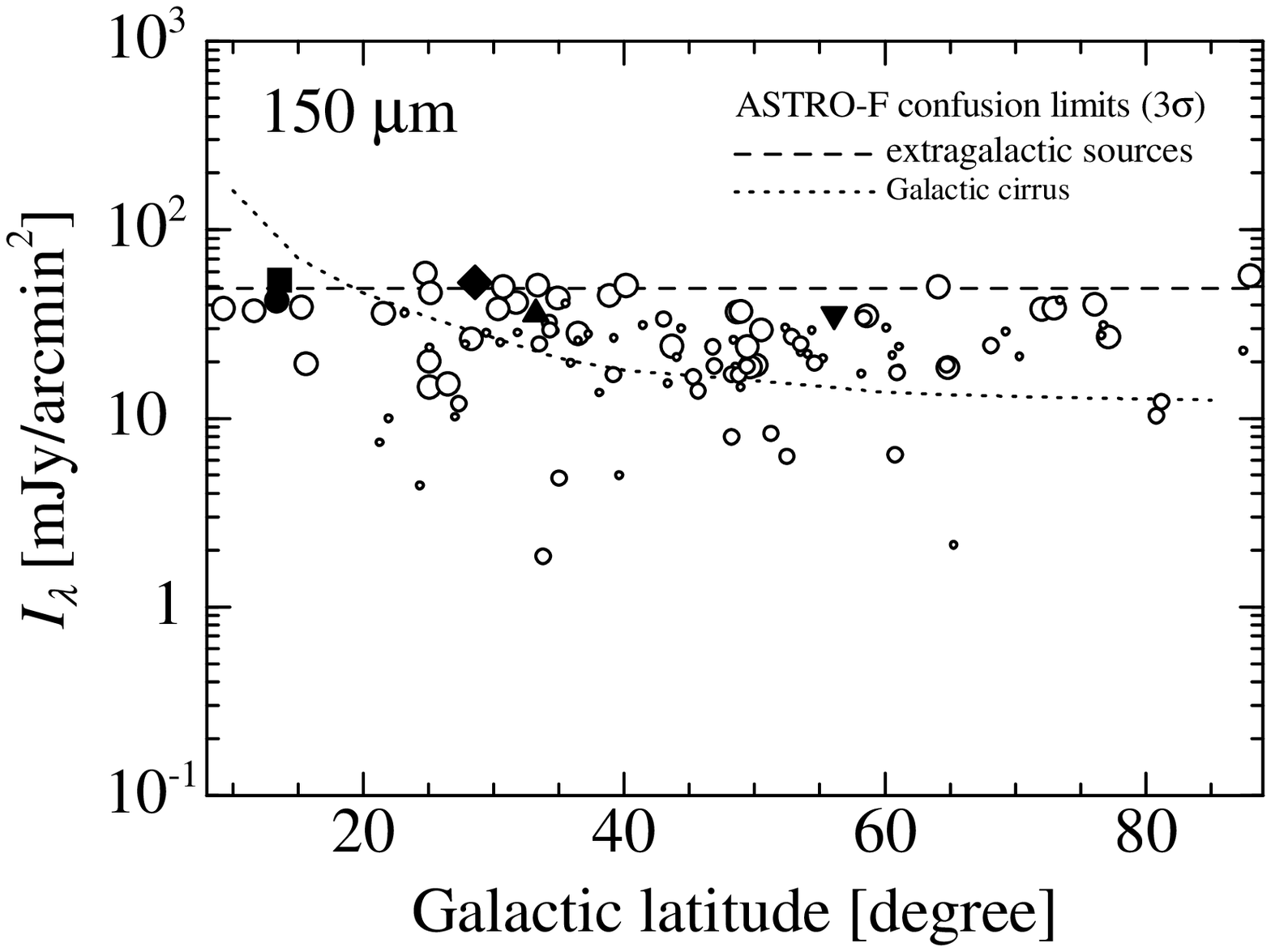}
    \FigureFile(84mm,60mm){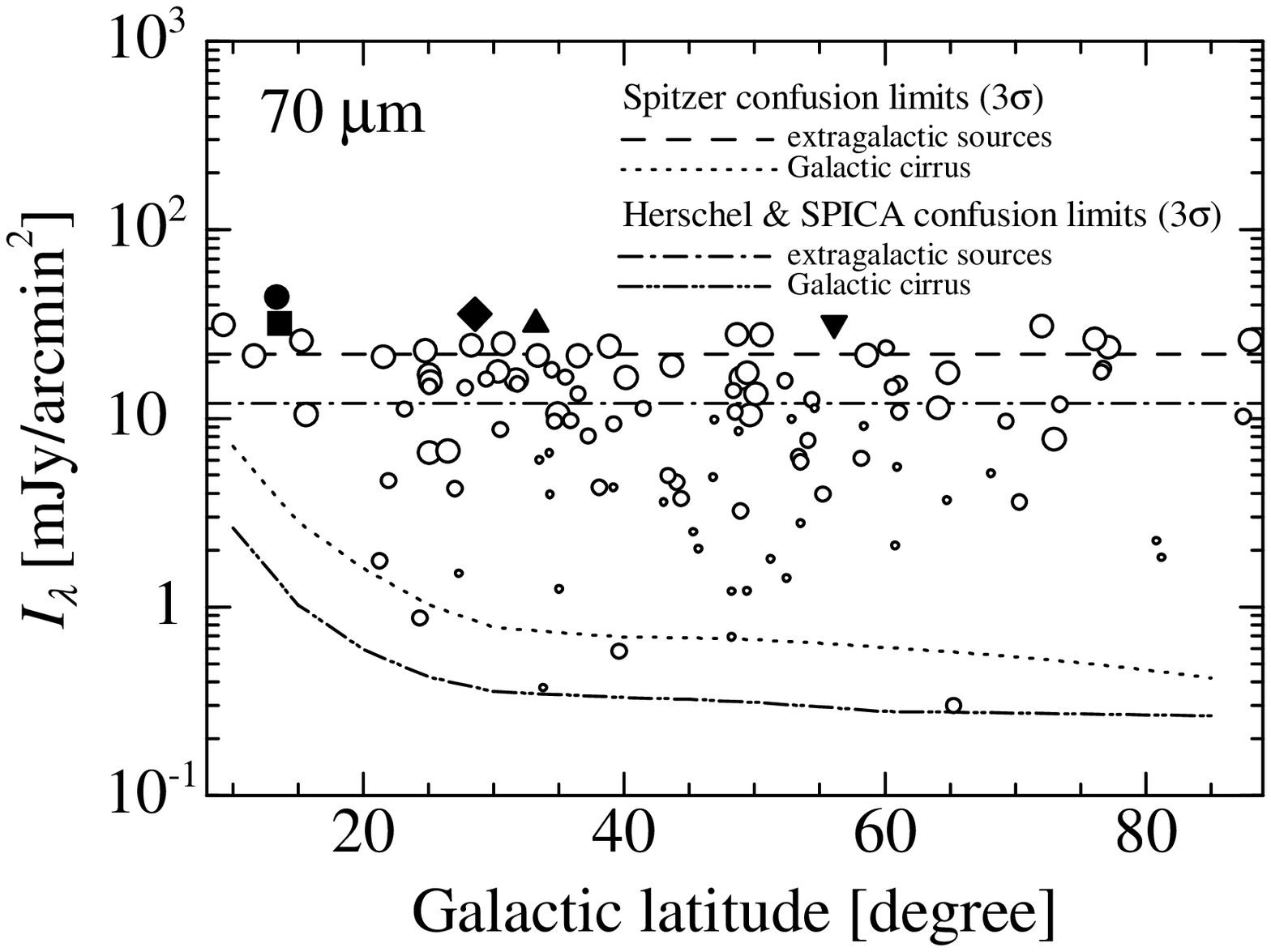}
    \FigureFile(84mm,60mm){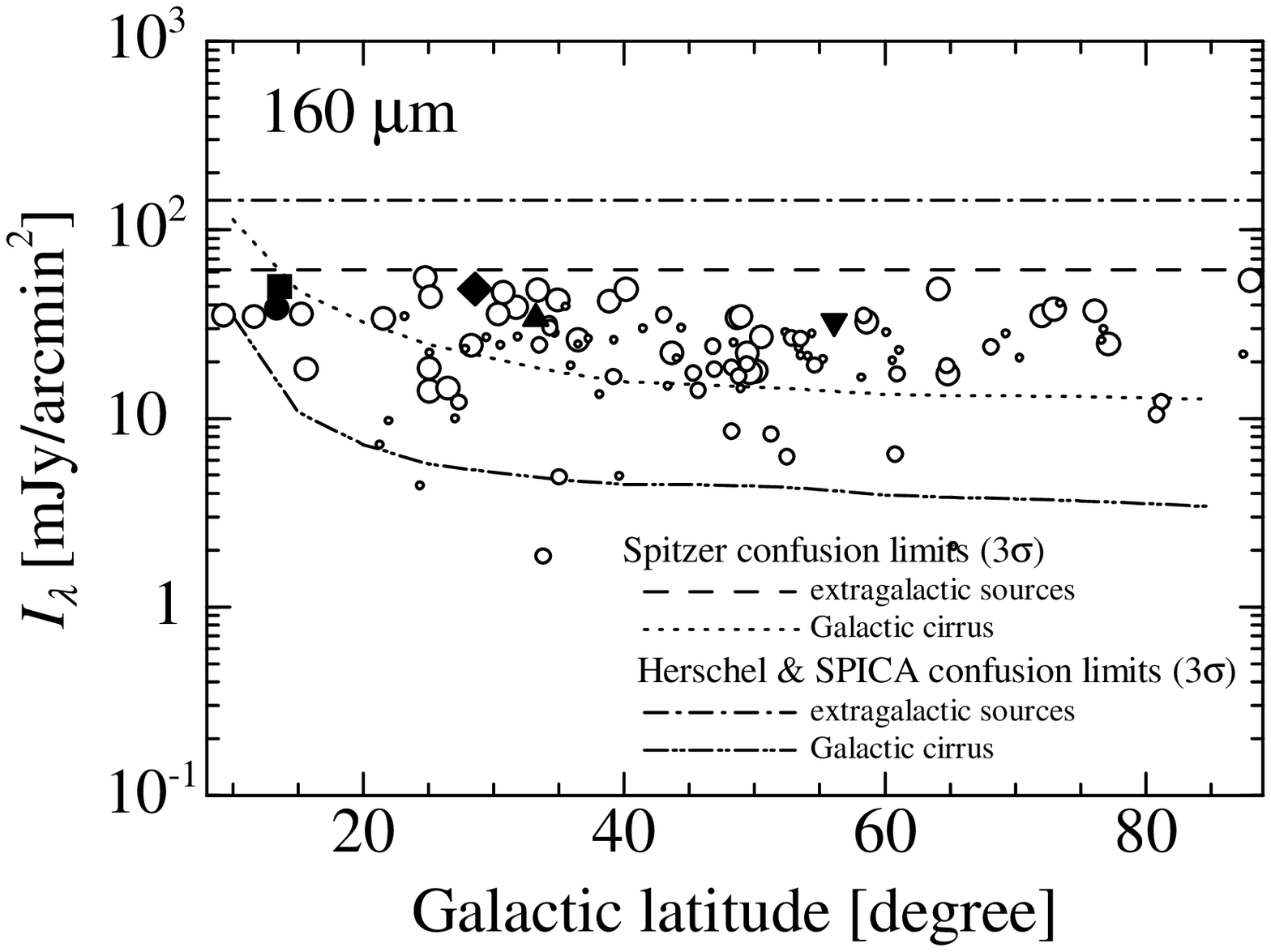}
    \FigureFile(84mm,60mm){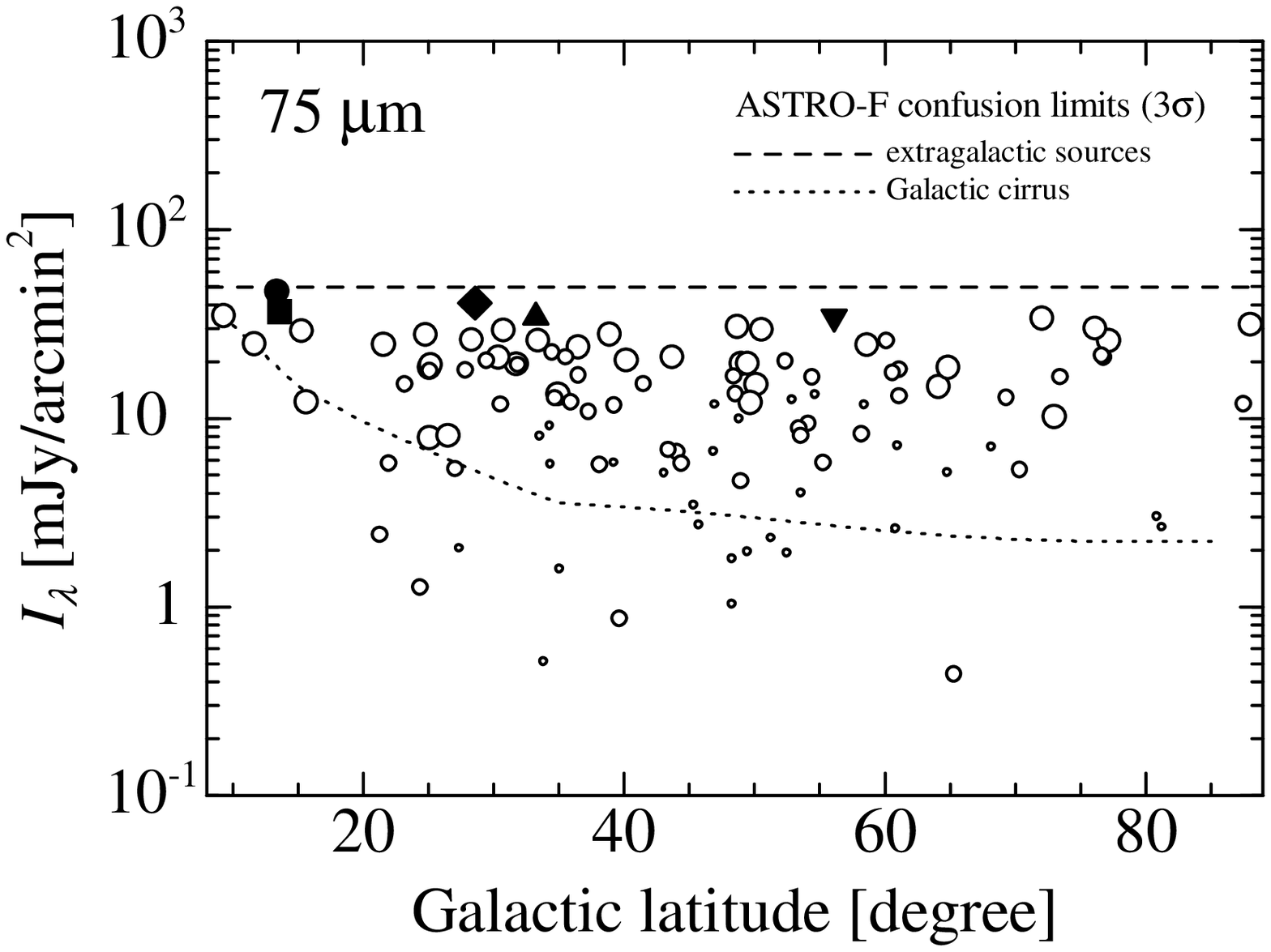}
    \FigureFile(84mm,60mm){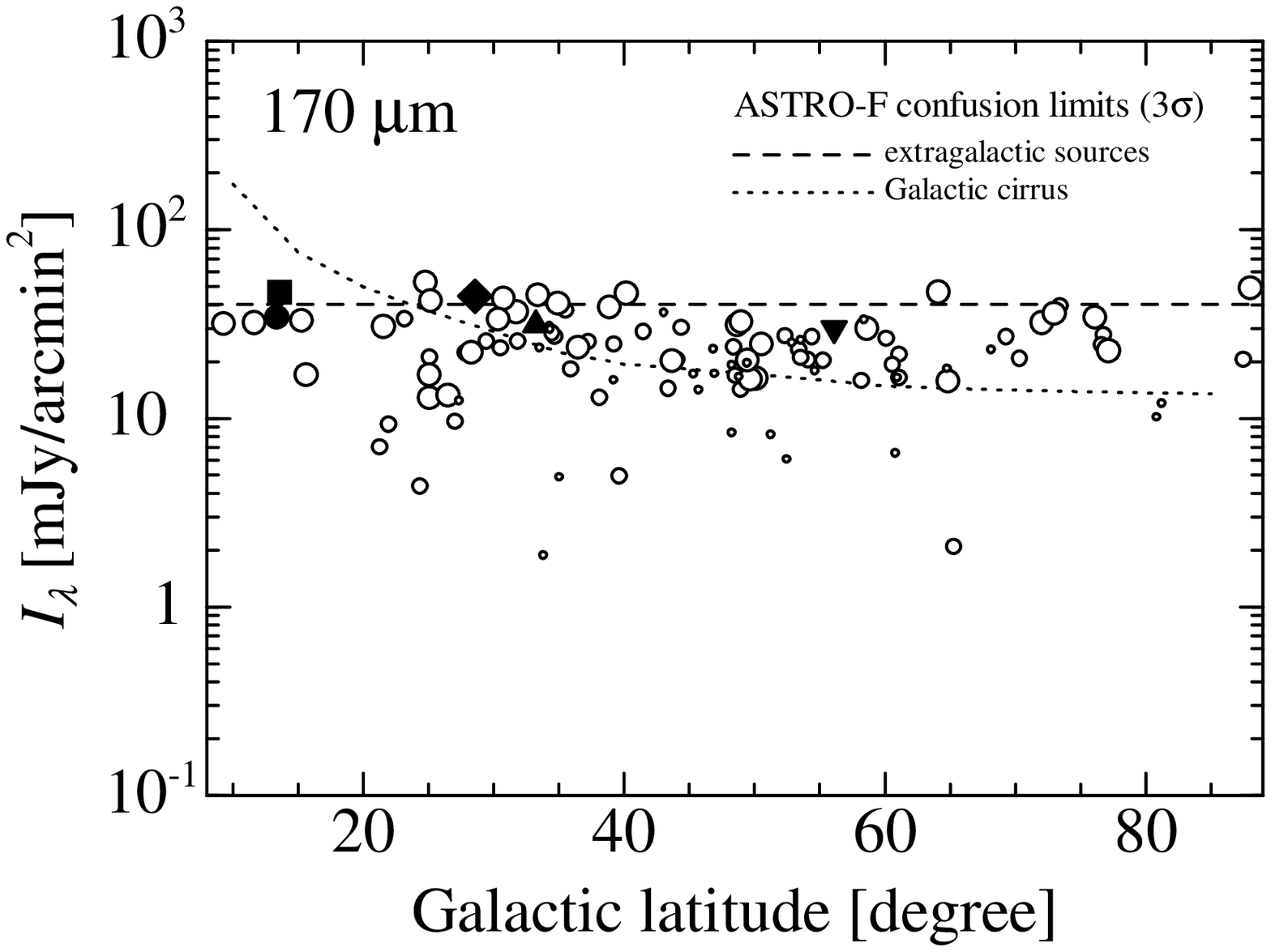}
  \end{center}
  \caption{Expected intensities at 60, 70, 75, 150, 160 and 170
  $\mu$m at the projected separation of 20 kpc from the center of 117
  galaxy clusters versus Galactic latitude. Five clusters with the
  highest 70 $\mu$m intensities are marked by filled symbols; Perseus
  (circle), A2319 (square), A3571 (lozenge), A2204 (triangle), and A3112 
  (reverse-triangle). The other clusters are marked by open circles;
  at redshift $0.01 < z < 0.1$ (large circles), at $0.1 \le z < 0.3$
  (medium circles), and at $0.3 \le z < 0.8$(small circles). Lines
  represent 3$\sigma$ confusion limits, due to extragalactic sources and
  Galactic cirrus for the mission indicated in each panel.} 
  \label{fig:Int_allcluster}
\end{figure*}

In Figure \ref{fig:Int_allcluster}, we plot the expected intensities of
the ICD for the entire sample of clusters as a function of the Galactic
latitude. In general, the highest signal-to-noise ratio (S/N) is
achieved at 70 $\mu$m. Five clusters with the highest 70 $\mu$m
intensity (filled symbols) all lie above the $4\sigma$ levels for
Spitzer and above the $7\sigma$ levels for Herschel and SPICA. The
emission tends to decrease toward high redshift because the observed
intensity and wavelength vary as $\propto (1 + z)^{-4}$ and $\propto 1 +
z$, respectively. Several nearby clusters at $z<0.1$ lie above the
$3\sigma$ levels of the detector at 70 $\mu$m, 150 $\mu$m, and 170
$\mu$m. They are likely to serve as plausible candidates for the
positive detection of the ICD.
  
We note that the confusion noise per beam, due to extragalactic
sources, drops slower than the size of the aperture squared at $\lambda
> 100$ $\mu$m (e.g., Dole et al. 2004). This makes the extragalactic
confusion per fixed sky area {\it increase} with an increasing telescope
size (Figs \ref{fig:Per_Int} and \ref{fig:Int_allcluster} ). At such
long wavelengths, one may be able to achieve higher S/N by adding
together the larger sky area, while high resolution detectors should
still be useful to exclude individual galaxies.

\begin{figure}
  \begin{center}
    \FigureFile(84mm,60mm){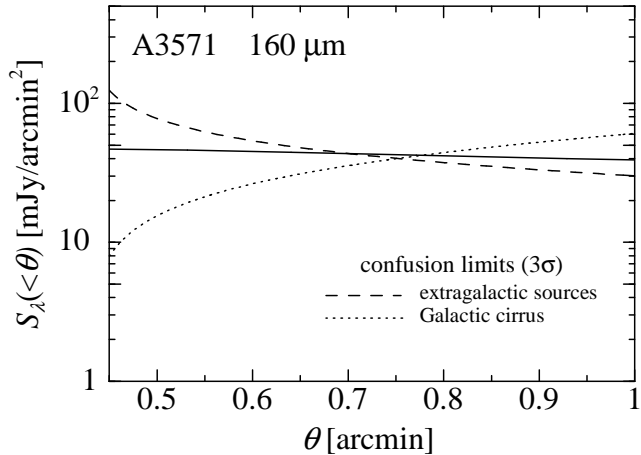}
  \end{center}
  \caption{Average surface brightness of A3571 at 160 $\mu$m 
  within a given radius excluding the central 20 kpc region. 
  Also plotted for reference are the confusion levels at 
  160 $\mu$m for the same area; due to extragalactic sources
  (dashed line) and Galactic cirrus (dotted line).}
  \label{fig:A3571_scaledep}
\end{figure}

In order to examine the scale dependence of the ICD detectability, we
plot in Figure \ref{fig:A3571_scaledep} the average surface brightness
of A3571 at 160 $\mu$m within a given radius excluding the central 20
kpc region. Also plotted for reference are the confusion levels for the
same area. We have chosen A3571 because it has the highest S/N at 160
$\mu$m among the five representative clusters mentioned above. In
computing the confusion noises for larger sky area, we have extrapolated
the values indicated in Table 1 and 2 of Dole et al. (2004)
for extragalactic sources and Figure 17 of Jeong et al. (2005)
for Galactic cirrus, respectively. The extragalactic noise
drops more rapidly with scale than the cluster signal, while the
Galactic confusion increases monotonically. The highest S/N is expected
within $\theta \sim 45$ arcsec. This sort of multi-scale analysis is a
powerful tool for detecting the ICD at $> 100$ $\mu$m. 

\subsection{Constraints on the ICD properties}
\label{sec:constraints}

We have so far assumed that the dust has been injected into the ICM with
essentially the same size distribution and amount as in our Galaxy. We
now examine how sensitive our results are to their uncertainties and
explore what constraints one can place in turn on the underlying dust
model from future observations.

Specifically, we vary three parameters in our model; the power-law index
$\alpha$ of the dust size distribution, their maximum size $a_{\rm
max}$, and the surface mass density $\Sigma_{\rm d}$ at the projected
radius $R=20$ kpc. Since very small dust grains are rapidly destroyed in
the ICM and do not contribute to mid- to far-infrared emissions, we
simply fix the minimum dust size at $a_{\rm min} = 0.001$ $\mu$m. As
mentioned in Section \ref{sec:model}, the power-law index $\alpha$ is
related to that of the {\rm injected} dust with $\alpha = \alpha^{\rm
inj} -1$ and has the value 2.5 for $\alpha^{\rm inj}=3.5$ (MRN). As
shown in Figure \ref{fig:Per_Sd}, $\Sigma_{\rm d}$ is nearly constant
in the cluster and serves as a good measure for the total amount of the
ICD. We take Perseus as a representative target and use the data at 24,
70 and 160 $\mu$m to derive constraints on the three parameters
mentioned above.

\begin{figure*}
  \begin{center}
    \FigureFile(84mm,60mm){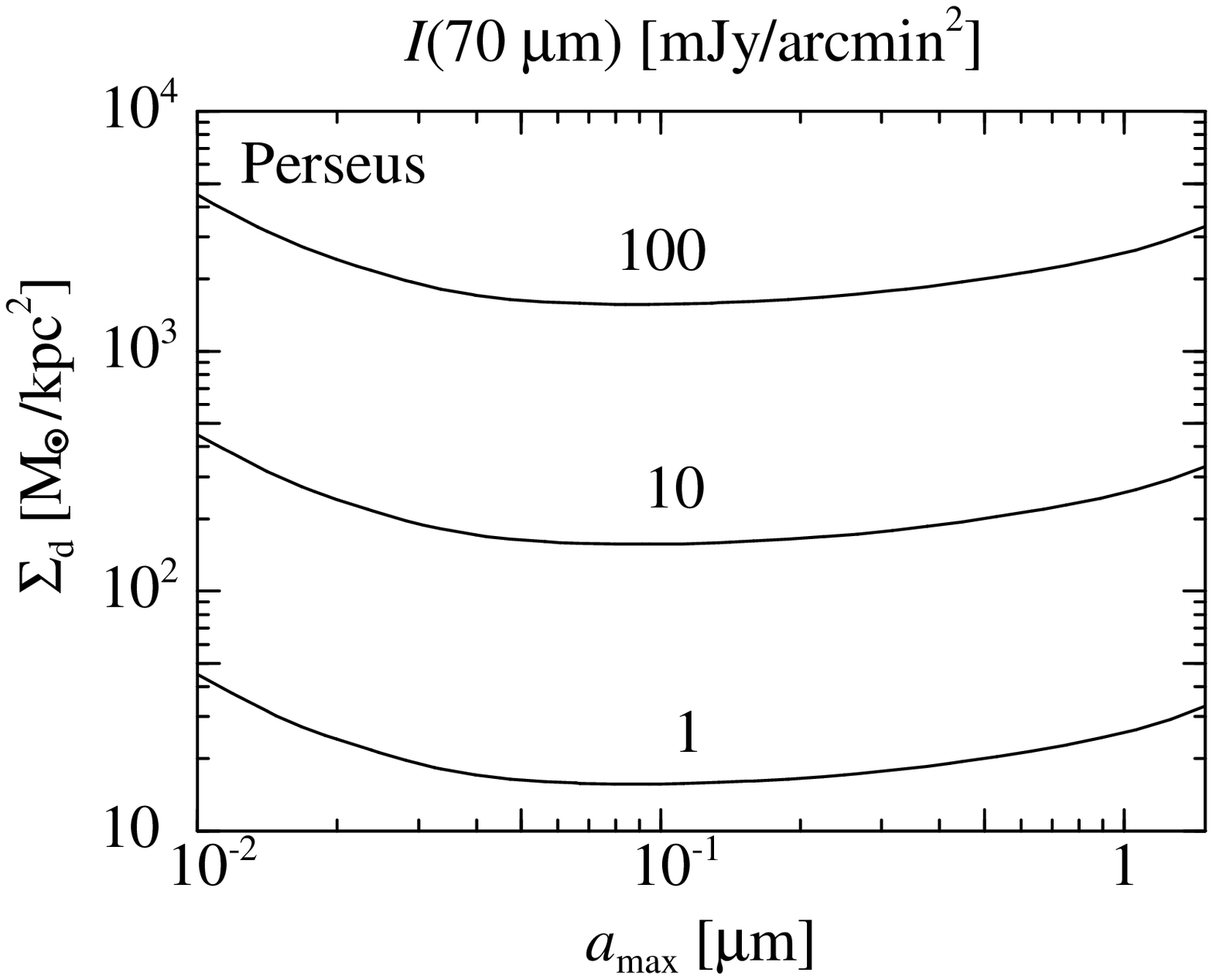}
    \FigureFile(84mm,60mm){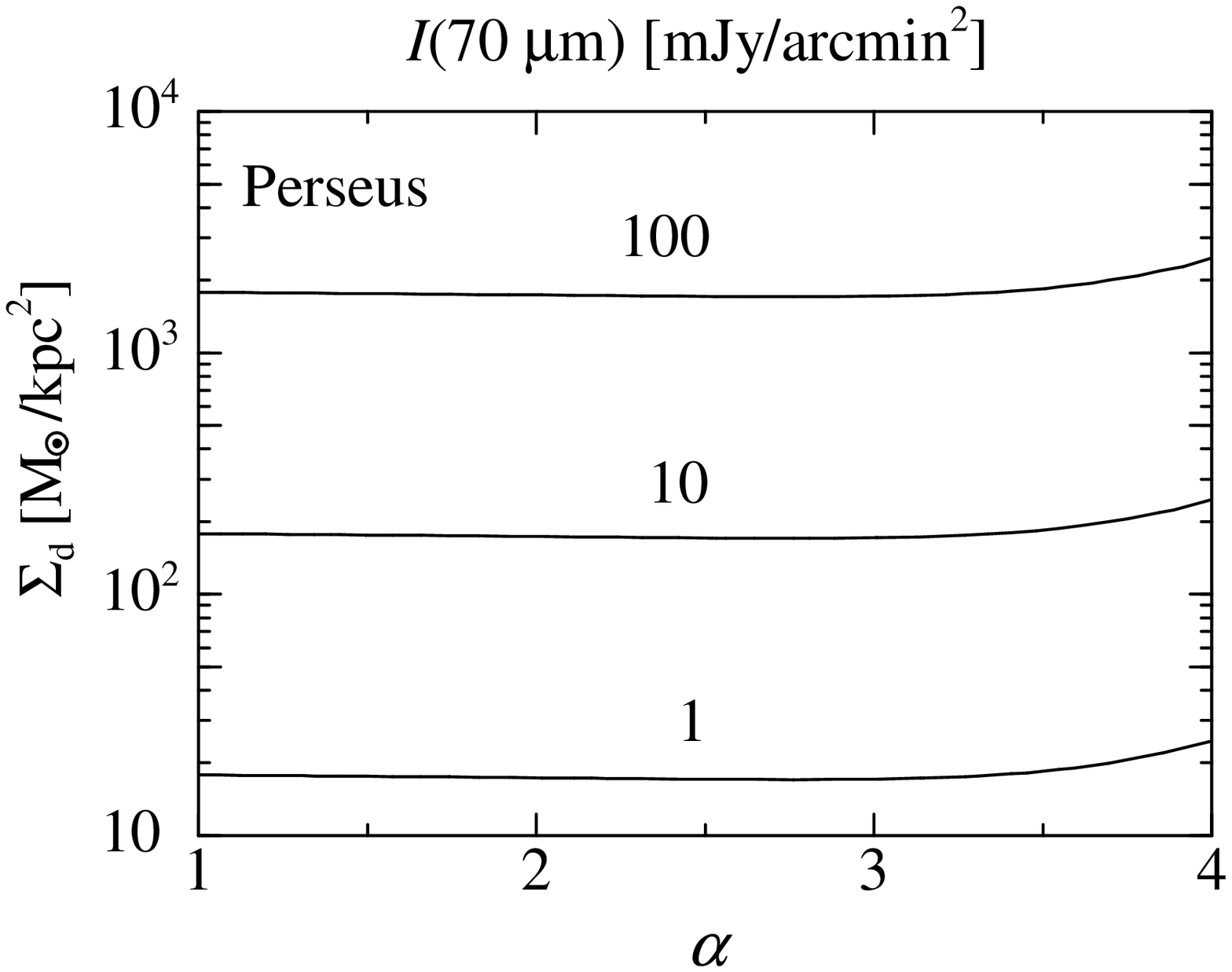}
  \end{center}
  \caption{Contours of the 70 $\mu$m intensity from Perseus on an 
  $a_{\rm max}-\Sigma_{\rm d}$ plane for $\alpha = 2.5$ (left panel) and
  an $\alpha-\Sigma_{\rm d}$ plane for $a_{\rm max} = 0.25 \mu$m (right
  panel). The curves  correspond to $I$(70 $\mu$m) = 
  100, 10 and 1 mJy/arcmin$^2$ from top to bottom in the both panels.}
 \label{fig:Per_I70Sdcont}
\end{figure*}

Figure \ref{fig:Per_I70Sdcont} shows contours of the 70 $\mu$m
intensity on $a_{\rm max}-\Sigma_{\rm d}$ plane for $\alpha = 2.5$ and 
$\alpha-\Sigma_{\rm d}$ plane for $a_{\rm max} = 0.25$ $\mu$m. We find 
that the intensity is much more sensitive to $\Sigma_{\rm d}$ than to 
$a_{\rm max}$ and $\alpha$. This indicates that the detection of 
(or an upper limit to) the ICD emission directly leads to a measure 
of the total amount of dust almost independently of its size 
distribution. Note that the mean ICD temperature varies only moderately 
and can be well constrained once electron density and temperature are 
known from X-ray observations (Fig. \ref{fig:Per_Tdave}).

\begin{figure*}
  \begin{center}
    \FigureFile(84mm,60mm){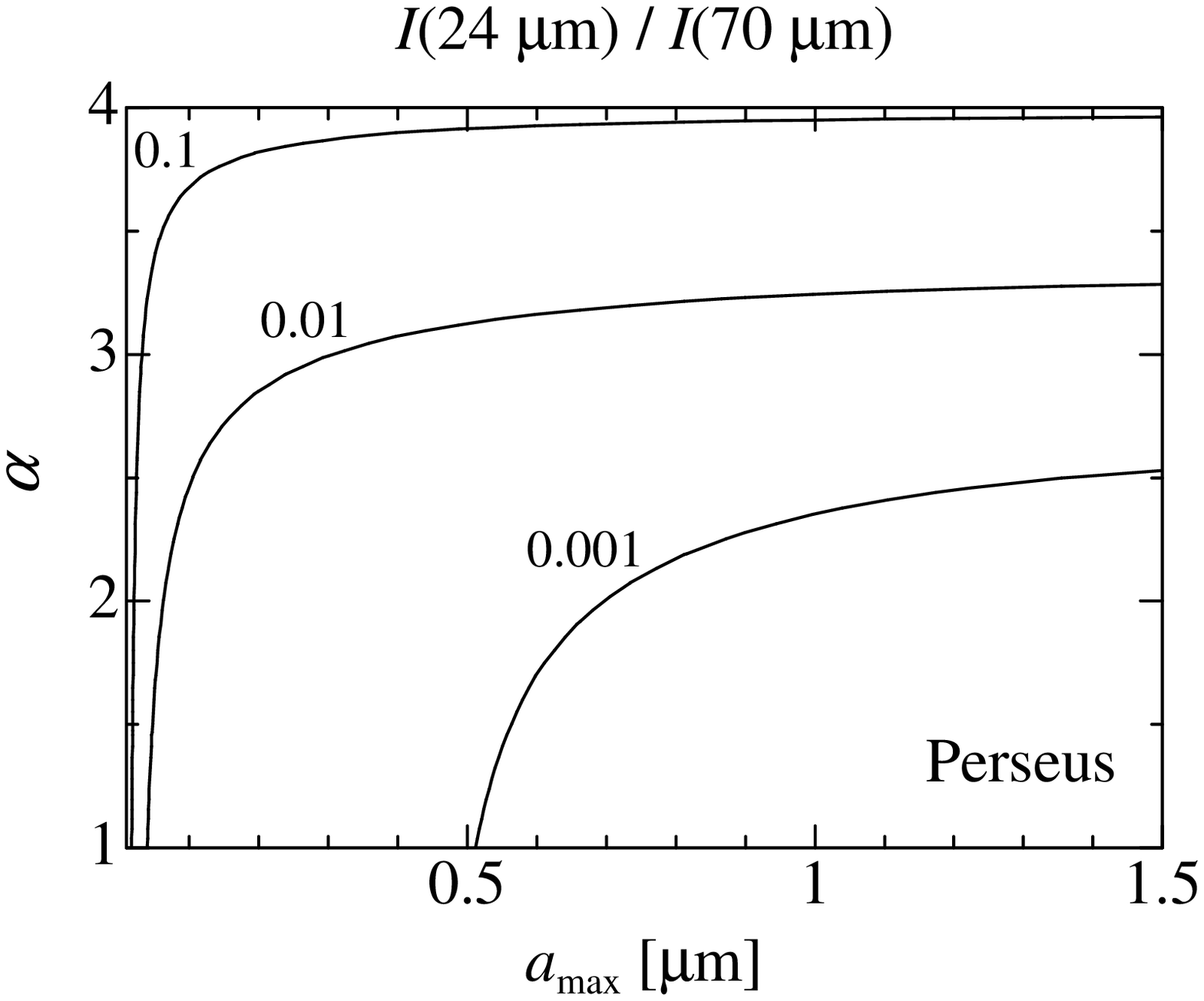}
    \FigureFile(84mm,60mm){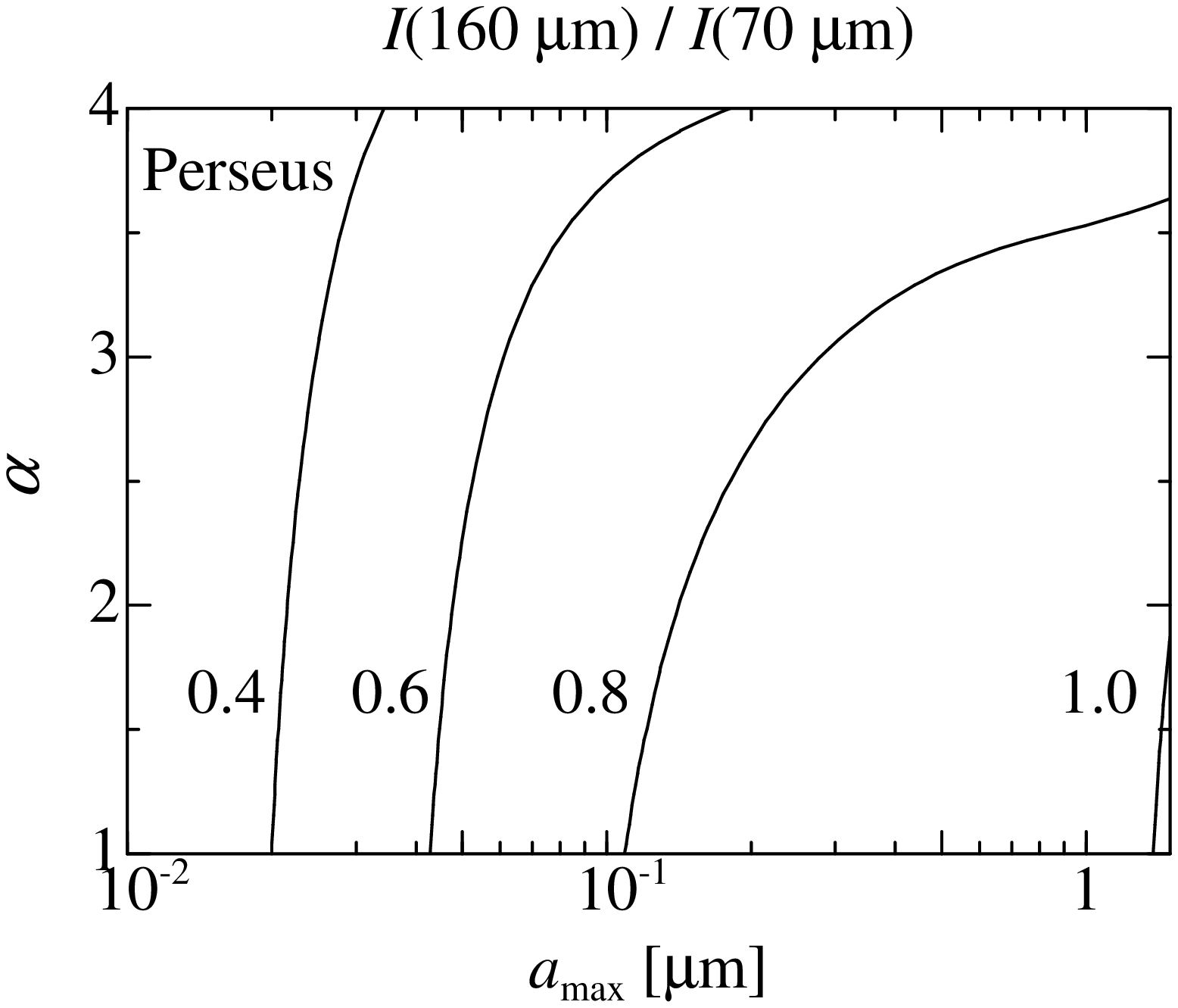}
  \end{center}
  \caption{Contours of the intensity ratios, $I$(24 $\mu$m)/$I$(70
 $\mu$m) (left panel) and $I$(160 $\mu$m)/$I$(70 $\mu$m) (right panel)
 on an $a_{\rm max}-\alpha$ plane for Perseus.  The curves, from top to
 bottom, correspond to $I$(24 $\mu$m)/$I$(70 $\mu$m) = 0.1, 0.01 and
 0.001 in the left panel and to $I$(160 $\mu$m)/$I$(70 $\mu$m) = 0.4,
 0.6, 0.8 and 1.0 in the right panel. } \label{fig:Per_Intratiocont}
\end{figure*}

Constraints on the size distribution, $a_{\rm max}$ and $\alpha$, will
be obtained via multi-band observations as illustrated in Figure
\ref{fig:Per_Intratiocont}. Since the intensities are 
proportional to $\Sigma_{\rm d}$, uncertainties in $\Sigma_{\rm
d}$ are eliminated by taking their ratios in multi-bands.  The ratio 
$I$(160 $\mu$m)/$I$(70 $\mu$m) can yield a reasonable measure of
$a_{\rm max}$ provided that $\alpha$ is not significantly different 
from that inferred in the MRN model $\alpha=2.5$. On the other hand, 
$I$(24$\mu$m)/$I$(70 $\mu$m) is more sensitive to $\alpha$ than 
$a_{\rm max}$ and can be used to infer the former quantity if 
sufficient S/N is achieved in both bands.  

\section{Discussion}
\label{sec:discussion}

There have been a number of suggestions as to how dust grains are
replenished into the intergalactic medium. For example, many authors 
have argued that a significant amount of dust could be expelled from 
galaxies by radiation pressure (e.g., Chiao \& Wickramasinghe 1972; 
Ferrara et al. 1991; Shustov \& Vibe 1995; Davies et al. 1998; Simonsen 
\& Hannestad 1999; Aguirre et al. 2001a,b). Radiation pressure may 
limit the injected dust size at $0.05-0.2$ $\mu$m (Shustov \& Vibe 1995
; Davies et al. 1998) because smaller grains are rapidly destroyed via 
sputtering in galactic halos while larger ones are hard to escape from 
the gravitational potential of galaxies. Ejection of small grains can 
be suppressed further by drag forces due to dust-gas collisions and 
Coulomb interactions (Bianchi \& Ferrara 2005). Ferrara et al. (1991) 
point out that radiation pressure may also change the composition of 
ejected grains because graphite grains can attain higher velocities 
and escape more easily from galaxies than silicate grains.

Another potential source of the ICD is from the intracluster stellar
population (Montier \& Giard 2004). Zwicky (1951) was the first to
suggest the presence of the intracluster stars based on the detection of
the excess light between the galaxies of the Coma cluster. Numerical
simulations indicate that $\sim10-20$ per cent of the total cluster
light should be the ICM (Murante et al. 2004; Willmanet et al. 2004). 
Recently, planetary nebula (e.g. Feldmeier et al. 2004; Theuns \& Warren
1997), red-giant-branch stars (Ferguson et al. 1998; Durrell et
al. 2002), and also Type Ia supernova (Gal-Yam et al. 2003) have been
detected in the ICM. The amount of intracluster light allowed from these
observations is $\sim10-50$ per cent of the total cluster
light. According to Montier \& Giard (2004), the dust-to-gas mass ratio
of the Coma-like cluster, due to the intracluster stellar population, is
$\sim 10^{-5}$ within a radius of 1 Mpc and predominant compared to the
galactic production. The intracluster stellar population can add small
dust and silicate grains that might be suppressed if galaxies are the
only source of the ICD. As demonstrated in Section \ref{sec:constraints},
multi-band observations of the dust emission will provide a powerful
probe of the grain size distribution, which can be used to constrain the
ejection processes of the ICD.

\section{Conclusions}\label{sec:conclusion}

In this paper, we have performed a comprehensive study on the nature of
diffuse dust in a sample of galaxy clusters at redshift $z \sim 0.01 -
0.8$.  Based upon recent X-ray data, the temperature distribution of the
dust grains are computed taking account of collisional heating by
ambient hot plasma and radiative cooling. The dust size distribution is
also solved by incorporating injection from galaxies and destruction via
sputtering.

If the dust grains are injected into the intergalactic medium with the
amount and size comparable to the Galactic values, the dust-to-gas ratio
is typically $\sim 10^{-6}$ and the mean dust temperature is $\sim 30$ K
near the cluster center. The predicted emissions lie marginally above
the detection thresholds for Spitzer Space Telescope, ASTRO-F, 
Herschel and SPICA missions. The highest signal-to-noise ratios 
($>4\sigma$) are expected for some nearby clusters such as Perseus, 
A3571, A2319, A3112 and A2204 at the 70 $\mu$m band.

Given rather tight constraints on the dust temperature from observed
electron density and temperature, the dust mass can be inferred directly
from the infrared observations. Further constraints on the size
distribution will be obtained once multi-band data are available by the
future facilities. They will definitely provide a powerful probe of the
dust injection processes and dust-gas interactions in the intergalactic
space.

\bigskip

We thank Woong-Seob Jeong for offering the confusion noise data due to
Galactic cirrus, Naomi Ota for providing the X-ray data of distant
clusters, and the referee Martin Giard for helpful comments. We also
thank Hidehiro Kaneda, Hirohisa Nagata, Mamoru Shimizu, and Hidenori
Takahashi for useful discussions. This work is supported in part by 
the Grants-in-Aid by the Ministry of Education, Science and Culture of 
Japan (14740133:TK).


\section*{References}
\small

\re Aguirre, A., Hernquist, L., Katz, N., Gardner, J., \& Weinberg, 
D. 2001a, ApJ, 556, L11 

\re Aguirre, A., Hernquist, L., Schaye, J., Katz, N., Weinberg, 
D. H., \& Gardner, J. 2001b, ApJ, 556, 521

\re Arnaud, K. A., \& Mushotzky, R. F. 1998, ApJ, 501, 119

\re Bianchi, S., \& Ferrara, A. 2005, MNRAS, 358, 379

\re Bullock, J. S., Kolatt, T. S., Sigad, Y., Somerville, R. S., 
Kravtsov, A. V., Klypin, A. A., Primack, J. R., \& Dekel, A. 2001, 
MNRAS 321, 559

\re Buote, D. A. 2000, MNRAS, 311, 176

\re Chiao, R. Y., \& Wickramasinghe, N. C. 1972, MNRAS, 159, 361

\re Davies, J. I., Alton, P., Bianchi, S., \& Trewhella, M. 1998, 
MNRAS, 300, 1006

\re Davis, D. S., Mulchaey, J. S., \& Mushotzky, R. F. 1999, ApJ,
 511, 34

\re Dole, H., Lagache, G., \& Puget, J. L. 2003, ApJ, 585, 617

\re Dole, H., et al. 2004, ApJS, 154, 93

\re Draine, B. T., \& Anderson, N. 1985, ApJ, 292, 494

\re Draine, B. T., \& Lee, H. M. 1984, ApJ, 285, 89 

\re Draine, B. T., \& Salpeter, E. E. 1979, ApJ, 231, 77

\re Durrell, P. R., Ciardullo, R., Feldmeier, J. J., Jacoby, G. H.,
 \& Sigurdsson S. 2002, ApJ, 570, 119

\re Dwek, E. 1986, ApJ, 302, 363

\re Dwek, E., Rephaeli, Y., \& Mather, J. C. 1990, ApJ, 350, 104

\re Feldmeier, J. J., Ciardullo, R., Jacoby, G. H.,
 \& Durrel, P. R. 2004 ApJ, 615, 196

\re Ferguson, H. C., Tanvir, N. R., \& von Hippel, T. 1998, 
Nature, 391, 461

\re Ferrara, A., Ferrini, F., Franco, J., \& Barsella, B. 
1991, ApJ, 381, 137

\re Gal-Yam, A., Maoz, D., Guhathakurta, P., \& Filippenko, A. V. 
2003, AJ, 125, 1087

\re Girardi, M., Mezzetti, M., Giuricin, G., \& Mardirossian, F. 
1992, ApJ, 394, 442

\re Goudfrooij, P., \& de Jong, T. 1995, ApJ, 298, 784

\re Hickson, P., Menon, T. K., Palumbo, G. G. C., \& 
Persic, M. 1989, ApJ, 341, 679

\re Jeong, W.-S., Lee, H. M., Pak, S., Nakagawa, T., Kwon, S. M., 
Pearson, C. P., \& White, G. J. 2005, MNRAS, 357, 535

\re Karachentsev, I. D., \& Lipovetskii, V. A. 1969, 
Soviet Phys., 12, 909

\re Kitayama, T., Komatsu, E., Ota, N., Kuwabara, T., Suto, Y., 
Yoshikawa, K., Hattori, M., \& Matsuo, H. 2004, PASJ, 56, 17

\re Komatsu, E., Kitayama, T., Suto, Y., Hattori, M., Kawabe, 
R., Matsuo, H., Schindler, S., \& Yoshikawa K. 1999, ApJ, 516, L1

\re Laor, A., \& Draine, B. T. 1993, ApJ, 402, 441

\re Li, A., \& Draine, B. T. 2001, ApJ, 554, 778

\re Maoz, D. 1995, ApJ, 455, L115 

\re Mathis, J. S., Rumpl, W., \& Nordsieck, K. H. 1977, ApJ, 217, 425

\re Mohr, J. J., Mathiesen, B., \& Evrard, A. E. 1999, ApJ, 517, 627

\re Montier, L. A., \& Giard, M. 2004, A\&A 417, 401

\re Montier, L. A., \& Giard, M. 2005, accepted for publication in A\&A

\re Murante G., et al. 2004, ApJ, 607, L83

\re Mushotzky, R., Loewenstein, M., Arnaud, K. A., Tamura, T., Fukazawa, 
Y., Mastushita, K., Kikuchi, K., \& Hatsukade, I. 1996, ApJ, 466, 686

\re Navarro, J. F., Frenk, C. S., \& White, S. D. M. 1996, ApJ, 462, 563

\re Ota, N., \& Mitsuda, K. 2004, A\&A 428, 757

\re Pearson, C. P., et al. 2004, MNRAS, 347, 1113

\re Renzini, A. 1997, ApJ, 488, 35

\re Shimizu, M., Kitayama, T., Sasaki, S., \& Suto Y. 2003, ApJ, 590, 197

\re Shustov, B. M., \& Vibe, D. Z. 1995, Astronomy Reports, 39, 578

\re Simonsen, J. T., \& Hannestad, S. 1999, A\&A, 351, 1 

\re Stickel, M., Lemke, D., Mattila, K., Haikala, L. K., \& Haas, M. 1998, 
A\&A, 329, 55

\re Stickel, M., Klaas, U., Lemke, D., \& Mattila, K. 2002, A\&A, 383, 367

\re Sulentic, J. W., \& De Mello Rabaca, D. F. 1993, ApJ, 410, 520 

\re Theuns, T., \& Warren, S. J. 1997, MNRAS, 284, L11

\re Tsai, J. C., \& Mathews, W. G. 1995, ApJ, 448, 84

\re White, D. A. 2000, MNRAS, 312, 663 

\re Willmanet, B., Governato, F., Wadsley, J., \& Quinn T. 2004, MNRAS, 
355, 159

\re Yoshikawa, K., \& Suto, Y. 1999, ApJ, 513, 549

\re Zwicky, F. 1951, PASP, 63, 61

\re Zwicky, F. 1962, in Problems in Extragalactic Research, ed. 
G. C. McVittie(New York:Macmillan), 149

\end{document}